# Effective dislocation lines in continuously dislocated crystals
# II. Congruences of effective dislocations

ANDRZEJ TRZĘSOWSKI

**Institute of Fundamental Technological Research,**
**Polish Academy of Sciences**
Świętokrzyska 21, 00-049 Warsaw, Poland
e-mail: atrzes@ippt.gov.pl or atrzes@wp.pl

The notion of a congruence of effective dislocation lines endowed with the nonvanishing local Burgers vector is introduced. Particularly, the class of congrunces of principal Volterra-type effective dislocation lines associated with the dislocation densities (tensorial as well as scalar) is distinguished in order to investigate the geometry of continuized defective crystals in terms of these densities. It is shown that effective dislocation lines can be endowed with the dislocation line tension and with a finite self-energy.

## 1. Introduction

The straight *edge dislocation* has a rigorously defined plane in which it can move. The plane, called a *slip plane*, includes the dislocation and its Burgers vector and the dislocation motion (in the Burgers vector direction) is called then a *glide motion*. Likewise, a curved edge dislocation has a rigorously defined *slip surface* called also its *glide surface*. The edge dislocation is then the so-called *prismatic dislocation* and the slip on this surface is called a *prismatic slip* [1]. The planes tangent to the slip surface of a prismatic dislocation are *local slip planes*. For a straight *screw dislocation* its Burgers vector is parallel to the dislocation line.

At low temperatures when the diffusion is difficult, and in the absence of a non-equilibrium concentration of point defects, the movement of dislocations is restricted entirely to glide. However, at higher temperatures an edge dislocation can move out of the glide surface normal to the Burgers vector by a process called its *climb* [2]. The glide and climb motions are two basic types of dislocation movements [2]. For example, a prismatic edge dislocation loop can move only by glide on a cylindrical surface, and if the loop expands or shrinks, climb must be occurring. There are also prismatic dislocations in the of forms of *cylindrical helices*. Namely, dislocations in the of forms of a long spiral have been observed in crystals [2]. The spiral dislocation lies on a cylinder whose axis is parallel to the Burgers vector, and the dislocation can glide on this cylinder. Consequently, the prismatic helical dislocation is *mixed* (that is it has the edge and screw components).

The *slip*, which is the most common manifestation of plastic deformation in crystalline solids, can be envisaged as sliding or successive displacement of one plane of atoms over another, on a distinguished slip plane (local or global). Discrete blocks of crystal between two slip planes remain undistorted [2]. Consequently, any *dislocation line* in the crystal can be treated as a line formed by means of a slip (homogeneous or not), such that the dislocation becomes a boundary between the slipped and unslipped parts of the crystal (e.g. [2], [3]). The *slip direction* is then parallel to the Burgers vector of the dislocation, and the *slip magnitude* equals the strength of dislocation (defined as the modulus of its Burgers vector). The above representation of a dislocation concerns flat as well as spatial dislocation lines [3] and the dislocations so represented are called *Volterra dislocations* [2].

On the other hand, it is known that the glide motion of many dislocations results in slip, and it is observed that globally (i.e. on a macroscale) this motion is accompa-nied by the occurrence of *slip surfaces* [2] in which Volterra dislocations can move. It is because, among others, the appearance of many dislocations generates a bend of originally straight lattice lines and consequently *crystal surfaces* being slip surfaces occur. Such crystal surfaces are called *glide surfaces* of Volterra dislocations. For example, in the case of the so-called *single glide* (in which the crystal deforms by a slip on one set of parallel crystal planes only), lattice lines originally normal to the plane of slip, form a *normal congruence*, i.e., the lines of the congruence are orthogonal trajectories of a family of *crystal surfaces* of the continuously dislocated Bravais crystal [4], [5].

We see that, in a continuized crystal with many dislocations [5], the line being a boundary between slipped and unslipped parts of the crystal and located on a slip surface can be distinguished. This line can be endowed, in the framework of the geometrical theory of dislocations, with the Burgers vector [5] as well as with the so-called *local Burgers vector* tangent to the slip surface along the line everywhere (Section 2 and e.g. [6], [7]). The line endowed with a local Burgers vector nonvanishing everywhere is called a *Volterra-type effective dislocation line* (Sections 3 and 4). The glide motion of a Volterra-type effective dislocation can be considered as a *mesoscopic* elementary act of macroplasticity. More generally, we can extend this definition of *effective dislocations* on each smooth curve (flat or spatial) that can be endowed with the local Burgers vector nonvanishing everywhere. We consider also dislocation lines understood in this broader sense (Sections 2 and 3). It is shown that effective dislocation lines can be endowed with the dislocation line tension and with a finite self-energy (Section 7).

It is known that the occurrence of many dislocations in a crystalline solid is accompanied with the appearance of point defects created by the distribution of dislo-cations [8]. The influence of these *secondary point defects* on metric properties of a continuously dislocated Bravais crystal can be modeled by the assumption that the body under consideration is additionally endowed with such Riemannian *internal length measurement* that reduces to the Euclidean metric of the body if dislocations are absent (Section 2; see also [5] and [6]). The influence of secondary point defects on the slip phenomenon can be then taken into account by means of the treatment of congruences of effective dislocation lines as those located in the Riemannian material space, and slip and crystal surfaces can be represented by 2-dimensional submanifolds of this space (Sections 2, 5 and 6). Note, that the such surfaces can be, at least locally, observed in the Euclidean ambient space of the dislocated crystalline body. It is because these surfaces can be locally isometrically embedded into this Euclidean space [9]. For example, crystal planes can be represented in the material space by the so-called *totally geodesic surfaces* [10], being an evident Riemannian generalization of planes of Euclidean 3-space (Section 5).

## 2. Local Burgers vector

Let $B \subset E^3$ be a *solid body* identified with its distinguished spatial configuration being an open and contractible to a point subset of the three-dimensional Euclidean point space $E^3$ [11]. We will consider the curvilinear coordinate systems $X = (X^A)$ defined on an open subset $U \subset B$ and such that $[X^A] = cm$, $A = 1, 2, 3$, and we will denote $p = p(X) \in U$ iff $X = X(p) \in \mathbb{R}^3$. The material structure of the body under consideration is defined as a continuous limit approximation of a Bravais crystal with many dislocations (see [5], Introduction). A distinguished basis $\Phi = (\mathbf{E}_a;\ a=1,2,3)$ of the linear module $W(B)$ (see [5], Appendix), called further on a *Bravais moving frame*, is considered as the one defining a system of three independent congruences of *lattice lines* of the continuized crystal as well as scales of an *internal length measurement* along these lines (Section 1 and [5]). The condition that the bend of lattice lines due to dislocations (Section 1) is not generated by a global deformation of the body means that the so-called *object of anholonomity* $C_{ab}^c \in C^\infty(B)$ defined by [5]

(2.1) $$[\mathbf{E}_a, \mathbf{E}_b] = C_{ab}^c \mathbf{E}_c,$$

does not vanish. The *Bravais moving coframe* $\Phi^* = (E^a)$ dual to $\Phi$ is a basis of the linear module $W^*(B)$ dual to $W(B)$ and uniquely defined by the condition:

(2.2) $$\mathbf{E}_a = e_a^A \partial_A, \quad E^a = e^a_A dX^A \quad \Rightarrow \quad \langle E^a, \mathbf{E}_b \rangle = e^a_A e_b^A = \delta_b^a,$$

where, according to the assumed dimensional convention, we have:

(2.3) $$[\mathbf{E}_a] = [\partial_A] = \text{cm}^{-1}, \quad [E^a] = [dX^A] = \text{cm}.$$

Note that although translational symmetries of the crystal are lost in the above mentioned continuous limit approximation, the base vector fields of a Bravais moving frame can be considered as those that define scales of an internal length measurement along local crystallographic directions of the dislocated Bravais crystal (Section 1). Namely, we can define the following intrinsic *material metric tensor* $\mathbf{g}$ of an *internal length measurement* within the dislocated Bravais crystal [5]:

(2.4) $$\mathbf{g} = \mathbf{g}[\Phi] = \delta_{ab} E^a \otimes E^b = g_{AB} dX^A \otimes dX^B,$$
$$g_{AB} = \delta_{ab} e^a_A e^b_B, \quad [\mathbf{g}] = \text{cm}^2,$$

where Eqs. (2.2) and (2.3) were taken into account. It is a Riemannian model of the distortion of the globally Euclidean length measurement within an ideal crystal $B$ due to many dislocations. Since the Riemannian metric is locally Euclidean, therefore it is an internal length measurement, consistent with the observed phenomenon that dislocations have no influence on the local metric properties of the crystalline body.

Next, we can define a nondimensional tensorial representation $\mathbf{S}[\Phi]$ of anholonomity of $\Phi$. Namely, if $\Phi = (\mathbf{E}_a)$ is a Bravais moving frame and $\Phi^* = (\mathrm{E}^a)$ is the *Bravais moving coframe* dual to $\Phi$, then we define [5], [12]:

$$\mathbf{S}[\Phi] = d\mathrm{E}^a \otimes \mathbf{E}_a = S_{ab}{}^c \mathrm{E}^a \otimes \mathrm{E}^b \otimes \mathbf{E}_c,$$
(2.5)
$$S_{ab}{}^c = -\frac{1}{2} C_{ab}^c, \qquad [S_{ab}{}^c] = \mathrm{cm}^{-1}.$$

$\mathbf{S}[\Phi]$ characterizes the existence of many dislocations in this sense that

(2.6) $\qquad \mathbf{S}[\Phi] = \mathbf{0} \quad \text{iff} \quad \mathbf{E}_a = \partial/\partial \xi^a, \quad a = 1, 2, 3,$

where $\xi = (\xi^a), [\xi^a] = \mathrm{cm}$, is a coordinate system on $B$. Thus, the tensor field $\mathbf{S}[\Phi]$ defines a measure of the *long-range distortion* of the dislocated Bravais crystal due to a bend of originally straight lattice lines [5]. This long-range distortion of the dislocated Bravais crystal can be quantitatively measured by the so-called *Burgers vector* $\mathbf{b}[\gamma]$ corresponding to a closed smooth contour $\gamma$ (a *Burgers circuit*) in the considered defective crystalline solid body $B$ [5]:

(2.7) $\qquad \mathbf{b}[\gamma] = \mathrm{b}^a[\gamma] \mathbf{C}_a, \qquad \mathrm{b}^a[\gamma] = \varepsilon \oint_\gamma \mathrm{E}^a,$

where $\varepsilon = \pm 1$ defines the Burgers vector orientation, $[\mathrm{b}^a[\gamma]] = 1$, and $\mathbf{C}_a$, $a = 1, 2, 3$, is an orthonormal Cartesian base of the Euclidean vector space $\mathbf{E}^3$ of translations in $E^3$. It seems physically reasonable to take into account the influence of secondary point defects on a Burgers vector. It can be done e.g. if its components are computed by means of an internal length measurement dependent on these point defects [5].

Namely, let us consider a Burgers circuit $\gamma \subset B$ as the one located in the *Riemannian material space* $B_g = (B, \mathbf{g})$ where $\mathbf{g} = \mathbf{g}[\Phi]$ is the material metric tensor defined by Eq. (2.4). Next, let us identify the vector fields $\mathbf{E}_a$ with linear differential operators and let $\Phi^* = (E^a)$ denotes the triple of base 1-forms dual to the such understood $\Phi$ (see [5], Appendix). Then, the components $b^a[\gamma]$ of a Burgers vector $\mathbf{b}[\gamma]$, can be treated as functionals in the Riemannian space $B_g$ defining, according to Eq. (2.7), a mapping $\gamma \subset B_g \to \mathbf{b}[\gamma] \in \mathbf{E}^3$. Let $\Sigma \subset B$ be a surface possessing the closed contour $\gamma$ as its boundary and treated as a two-dimensional compact, connected and oriented Riemannian submanifold of $B_g$. Since

$$\text{(2.8)} \qquad b^a[\gamma] = \varepsilon \int_\Sigma dE^a,$$

the Stokes theorem in a Riemannian manifold (e.g. [13], [14]) states that the components $b^a[\gamma]$ can be written in the following form [5]:

$$\text{(2.9)} \qquad \begin{array}{ll} b^a[\gamma] = \int_\Sigma \alpha^{ba} d\Sigma_b, & d\Sigma_b = d\Sigma l_b, \\ l_a = \delta_{ab} l^b, & \|\mathbf{l}\|_g = l_a l^a = 1, \qquad \mathbf{l} = l^a \mathbf{E}_a, \end{array}$$

where $\mathbf{l} \in W(B)$ is an unit vector field normal to the surface element $d\Sigma$ of $\Sigma$, and

$$\text{(2.10)} \qquad \alpha^{ba} = \varepsilon S_{cd}{}^a e^{cdb}, \qquad [\alpha^{ba}] = \text{cm}^{-1},$$

where and $e^{abc} \doteq \varepsilon^{abc}$ denotes the permutation symbol associated with the Bravais moving frame $\Phi = (\mathbf{E}_a)$ and considered as components in the base $\Phi$ of a contravariant 3-vector density of weight +1 in $B_g$. As a rule, the case $\varepsilon = -1$ is considered in the literature and then the tensor field

$$\text{(2.11)} \qquad \boldsymbol{\alpha} = \alpha^{ab} \mathbf{E}_a \otimes \mathbf{E}_b \qquad [\boldsymbol{\alpha}] = \text{cm}^{-3},$$

is called the *dislocation density tensor* or the *Ney's tensor*. We see that it is a tensor defined up to the choice of the Burgers vector orientation. Likewise, the *scalar volume dislocation density* $\rho$ of a finite total length $L_d(B)$ of dislocation lines located in $B$ will be measured with respect to the *material volume element* $dV_g$ of $B_g$:

$$0 < L_d(B) = \int_B \rho \omega_g = \int_B \rho dV_g < \infty, \quad [L_d(B)] = \text{cm}, \quad [\rho] = \text{cm}^{-2},$$

(2.12) $\quad \omega_g = E^1 \wedge E^2 \wedge E^3 = e\, dX^1 \wedge dX^2 \wedge dE^3, \quad dV_g = \sqrt{g}\, dX^1 dX^2 dX^3,$

$$e = \det\left(e^a_A\right) = \sqrt{g}, \quad g = \det(g_{AB}), \quad [\omega_g] = \text{cm}^3,$$

where $\omega_g$ is the volume 3-form and Eqs. (2.2) and (2.4) were taken into account.

Let us rewrite Eq. (2.9) in the following form:

(2.13)
$$b^a[\gamma] = \int_\Sigma \beta^a d\Sigma,$$
$$\beta^a = l_b \alpha^{ba}, \quad [\beta^a] = \text{cm}^{-1}.$$

Since the quantity $\delta b^a = \beta^a d\Sigma$ has the same dimension as the components of a Burgers vector have, the object $\delta \mathbf{b} = \delta b^a \mathbf{C}_a$ is usually considered as a continuum local version of this vector [15] (see also e.g.[16]). We will not consider this infinitesimal continuum version of the Burgers vector but we will define the *local Burgers vector* as a vector field $\mathbf{b}$ tangent to $B_g$ and such that [1]

(2.14)
$$\mathbf{b} = b^a \mathbf{E}_a, \quad \rho b^a = \beta^a, \quad [b^a] = \text{cm}, \quad [\mathbf{b}] = 1;$$
$$b_g = \|\mathbf{b}\|_g = \left(b^a b_a\right)^{1/2} > 0, \quad b_a = \delta_{ac} b^c, \quad [b_g] = \text{cm},$$

or equivalently:

(2.15) $\quad \rho \mathbf{b} = \mathbf{l}\alpha, \quad b_g = \|\mathbf{b}\|_g > 0.$

Let us denote by $C[\mathbf{l}]$ a congruence in the material space $B_g$ defined by such the unit vector field $\mathbf{l}$ that the local Burgers vector $\mathbf{b}$ of Eq. (2.15) is well-defined. We will identify a geometric curve of this congruence (see Appendix) with an *effective* *dislocation line* (see Section 1). This effective dislocation line located in $B_g$ can be interpreted as the *edge* (effective) dislocation line if [18]

(2.16) $$\mathbf{b} \cdot \mathbf{l} = b^a l_a = b_g m^a l_a = 0,$$

or - as the *screw* (effective) dislocation line if

(2.17) $$\mathbf{b} = \eta \mathbf{l}, \quad \eta \neq 0, \quad [\eta] = \text{cm}.$$

In other cases the effective dislocation line is called *mixed*.

The components $\alpha^{ab}$ of the dislocation density tensor can be written in the following form [5]:

(2.18) $$\alpha^{ab} = \gamma^{ab} + \sigma^{ab},$$
$$\gamma^{ab} = \alpha^{(ab)}, \quad \sigma^{ab} = \alpha^{[ab]} = \frac{1}{2} t_c e^{cab},$$

where, according to Eqs. (2.5) and (2.10), we have:

(2.19) $$t_a = e_{abc} \alpha^{bc} = \varepsilon C^b_{ab}, \quad \varepsilon = \pm 1$$

and $e_{abc} \doteq \varepsilon_{abc} (= \varepsilon^{abc})$ denotes the permutation symbol associated with the Bravais moving coframe $\Phi^* = (E^a)$ and considered as components in the base $\Phi^*$ of a covariant 3-vector density of weight $-1$ in $B_g$. It follows from Eqs. (2.5), (2.10), (2.18) and (2.19) that

(2.20) $$\varepsilon C^c_{ab} = t_{[a} \delta^c_{b]} - e_{abd} \gamma^{dc}.$$

Therefore, the long-range distortion of the continuously dislocated Bravais crystal with secondary point defects characterizes the pair $(\gamma, \mathbf{t})$, where [5]

(2.21) $$\gamma = \gamma^{ab} \mathbf{E}_a \otimes \mathbf{E}_b, \quad \gamma^{ab} = \gamma^{ba},$$
$$\mathbf{t} = t^a \mathbf{E}_a, \quad t^a = \delta^{ab} t_b; \quad [\gamma^{ab}] = [t^a] = \text{cm}^{-1}.$$

Introducing designations:

(2.22)
$$\cos\varphi_{l,t} = \frac{l \cdot t}{t_g}, \qquad 0 \leq \varphi_{l,t} \leq \pi,$$
$$t_g = \|\mathbf{t}\|_g = \left(t^a t_a\right)^{1/2}, \qquad [t_g] = \mathrm{cm}^{-1},$$

and

(2.23)
$$\boldsymbol{\mu} = \mu^a \mathbf{E}_a = \mu \mathbf{m}, \qquad \|\mathbf{m}\|_g = 1,$$
$$\mu^a = \frac{1}{2} t_b l_c e^{bca}, \qquad \mu = \frac{1}{2} t_g \sin\varphi_{l,t} \geq 0,$$

we can write, according to Eqs. (2.13)-(2.15), (2.18), (2.19) and (2.21)-(2.23), the local Burgers vector **b** in the form

(2.24)
$$\rho \mathbf{b} = \gamma \mathbf{l} + \mu \mathbf{m}, \qquad \mu \geq 0,$$
$$\|\mathbf{l}\|_g = \|\mathbf{m}\|_g = 1, \qquad \mathbf{l} \cdot \mathbf{m} = \mathbf{t} \cdot \mathbf{m} = 0.$$

It follows from Eq. (2.24) that

(2.25)
$$\rho \mathbf{b} \cdot \mathbf{l} = |\gamma|.$$

Thus, according to Eqs. (2.15) and (2.25), a dislocation line of $C[\mathbf{l}]$ is *mixed* iff

(2.26)
$$|\gamma| = \rho b_g \cos\varphi_{b,l},$$
$$\cos\varphi_{b,l} = \frac{\mathbf{b} \cdot \mathbf{l}}{b_g}, \qquad b_g = \|\mathbf{b}\|_g > 0.$$

Particularly, it is the *edge* dislocation line iff

(2.27)
$$|\gamma| = 0,$$

or it is the *screw* dislocation line iff

(2.28)
$$|\gamma| = \rho \eta, \qquad b_g = |\eta| > 0.$$

The local movement of an effective edge dislocation line is limited to a specific local plane. Although the local movement of an effective screw dislocation can also be envisaged to take place in a local slip plane, nevertheless the line of the screw dislocation and the local Burgers vector do not define a unique plane.

## 3. Principal congruences

A congruence of (effective) dislocation lines $C[\mathbf{l}]$ is called *principal* if $\mathbf{l}$ is an *eigenvector* of the tensor field $\boldsymbol{\gamma}$ of Eq. (2.21), i.e. ([7], [17]):

(3.1) $$\boldsymbol{\gamma}\mathbf{l} = \gamma\mathbf{l}, \qquad \|\mathbf{l}\|_g = 1, \qquad \gamma \in \mathbb{R}.$$

We will say then that $\mathbf{l}$ defines a *principal direction* of $\boldsymbol{\gamma}$ and that $C[\mathbf{l}]$ consists of *principal (effective) dislocation lines*. In this case

(3.2) $$\rho b_g = \sqrt{\gamma^2 + \mu^2} > 0.$$

If $\mathrm{rank}\,\boldsymbol{\gamma} \geq 2$, then the orthonormal triple $\Gamma = (\mathbf{l}_a;\, a=1,2,3)$ of eigenvectors of $\boldsymbol{\gamma}$ is defined uniquely up to its orientation and we have the following representation:

(3.3) $$\boldsymbol{\gamma} = \gamma^a \mathbf{l}_a \otimes \mathbf{l}_a, \qquad \gamma^a \in C^\infty(B),$$
$$\mathbf{l}_a \cdot \mathbf{l}_b = \delta_{ab}, \qquad [\mathbf{l}_a] = [\gamma^a] = \mathrm{cm}^{-1},$$

where, according to Eqs. (2.2) and (2.4), should be:

(3.4) $$\mathbf{l}_a = Q^b{}_a \mathbf{E}_b, \qquad Q^b{}_a \in C^\infty(B),$$
$$\mathbf{Q} = (Q^a{}_b;\, {}^{a\downarrow 1,2,3}_{b\to 1,2,3}): B \to SO(3).$$

Therefore,

(3.5) $$\boldsymbol{\gamma} = \mathbf{Q}\boldsymbol{\eta}\mathbf{Q}^T, \qquad \boldsymbol{\eta} = \gamma^a \mathbf{E}_a \otimes \mathbf{E}_a, \qquad \mathrm{rank}\,\boldsymbol{\eta} \geq 2$$

and, according to Eqs (2.22)-(2.24), the corresponding triple $B = (\mathbf{b}_a)$ of *principal local Burgers vectors* is given by:

(3.6) $$\rho \mathbf{b}_a = \gamma_a \mathbf{Q}\mathbf{E}_a + \mu_a \mathbf{m}_a, \qquad a = 1,2,3,$$
$$\mathbf{Q}: B \to SO(\mathbf{E}^3), \qquad \mu_a = \frac{1}{2} t_g \sin \varphi_{\mathbf{l}_a,\mathbf{t}} \geq 0,$$

where, according to Eqs. (2.25), (2.26) and (3.1)-(3.3), we have:

(3.7)
$$\rho b_{g,a} = \sqrt{\gamma_a^2 + \mu_a^2} > 0, \qquad \gamma_a = \delta_{ab}\gamma^b,$$
$$\cos\varphi_a = \gamma_a / \rho b_{g,a}, \qquad \varphi_a = \varphi_{\mathbf{b}_a,\mathbf{l}_a}.$$

Let us consider a particular case defined by the following condition:

(3.8)
$$\varphi_{\mathbf{l}_\alpha,\mathbf{t}} = \pi/2, \qquad \alpha = 1,2.$$

In this case, according to Eqs. (2.23) and (3.3), we have

(3.9)
$$\mathbf{t} = t_g \mathbf{l}_3, \qquad t_g \geq 0,$$

and

(3.10)
$$\rho\mathbf{b}_1 = \gamma^1 \mathbf{l}_1 + \mu \mathbf{m}_1, \qquad \rho\mathbf{b}_2 = \gamma^2 \mathbf{l}_2 + \mu \mathbf{m}_2,$$
$$\rho\mathbf{b}_3 = \gamma^3 \mathbf{l}_3, \qquad \mu = t_g/2$$

where, taking into account that $(\mathbf{l}_1, \mathbf{l}_2, \mathbf{l}_3)$ is an ordered triple of vector fields, we have:

(3.11)
$$\mathbf{m}_1 = \mathbf{l}_2, \qquad \mathbf{m}_2 = -\mathbf{l}_1.$$

It follows from Eqs. (3.7) and (3.8) that

(3.12)
$$\rho b_{g,\alpha} = \sqrt{(\gamma^\alpha)^2 + \mu^2},$$
$$\rho b_{g,3} = |\gamma^3|, \qquad b_{g,\alpha} = \|\mathbf{b}_\alpha\|, \qquad \alpha = 1,2,$$

and

(3.13)
$$\cos\psi = \frac{\mathbf{b}_1 \cdot \mathbf{b}_2}{b_{g,1} b_{g,2}} = \frac{t_g}{2\rho^2 b_{g,1} b_{g,2}} (\gamma^2 - \gamma^1).$$

Thus

(3.14)
$$\psi = \frac{\pi}{2} \quad \text{iff} \quad (\gamma^2 - \gamma^1) t_g = 0.$$

If Eq. (3.8) and the condition

(3.15)
$$\gamma^3 = 0$$

are fulfilled, then $\mathbf{b}_3 = \mathbf{0}$ what means that the principal vector $\gamma_3$ of $\gamma$ does not define a congruence of effective dislocation lines. If additionally the 2-dimensional distribution $\pi(\mathbf{E}_1, \mathbf{E}_2)$ of planes (see [5], Appendix) consists of local crystal planes being virtually local slip planes, then we can take without loss of generality that

(3.16) $$\mathbf{l}_3 = \mathbf{E}_3.$$

The components $Q^a{}_b$ of Eq. (3.4) take then the following form:

(3.17) $$Q^a{}_b = \cos\phi\, \delta^a_b + (1-\cos\phi)\delta^a_3 \delta_{3b} - \sin\phi\, \varepsilon^a{}_{b3},$$
$$\varepsilon^a{}_{bc} = \delta^{ad}\varepsilon_{dbc}, \quad \phi: B \to \langle 0, \pi\rangle, \quad a, b = 1, 2, 3.$$

So, in this case [18]:

(3.18) $$\mathbf{l}_1 = \cos\phi\, \mathbf{E}_1 + \sin\phi\, \mathbf{E}_2,$$
$$\mathbf{l}_2 = -\sin\phi\, \mathbf{E}_1 + \cos\phi\, \mathbf{E}_2,$$

and

(3.19) $$\rho\mathbf{b}_1 = (\gamma^1 \cos\phi - \mu\sin\phi)\mathbf{E}_1 + (\gamma^1 \sin\phi + \mu\cos\phi)\mathbf{E}_2,$$
$$\rho\mathbf{b}_2 = -(\gamma^2 \sin\phi + \mu\cos\phi)\mathbf{E}_1 + (\gamma^2 \cos\phi - \mu\sin\phi)\mathbf{E}_2.$$

Note that Eq. (3.4) can be considered as a local rotational uncertainty to select the Bravais moving frame $\Phi$ [5]. Such uncertainty appears in the gauge theory of dislocations [11] and then Eq. (3.17) means the local transverse isotropy of the distortion of a crystalline body due to dislocations.

The principal Burgers vectors $\mathbf{b}_1$ and $\mathbf{b}_2$ of Eq. (3.19) are tangent to lattice lines, like it takes place for single dislocations in real crystal structures [2], if e.g.

(3.20) $$\operatorname{tg}\phi = \frac{\mu}{|\gamma|}, \quad \phi \in (0, \pi/2),$$

where

(3.21) $$\gamma^1 = \gamma^2 = \gamma \neq 0, \quad \mu = t_g/2 \geq 0.$$

Then

$$(3.22) \qquad \rho \mathbf{b}_\alpha = \frac{\gamma}{\cos\phi} \mathbf{E}_\alpha, \qquad \gamma \neq 0; \qquad \alpha = 1, 2,$$

and it follows from Eqs. (3.12), (3.21), and (3.22) that

$$(3.23) \qquad \mathbf{b}_1 \cdot \mathbf{b}_2 = 0, \qquad b_{g,1} = b_{g,2} = b_g,$$

where

$$(3.24) \qquad \rho b_g = |\gamma|/\cos\varphi > 0.$$

If $t_g = 0$ (that is $\phi = 0$) then, according to Eqs. (3.18) and (3.22), we are dealing with principal effective *screw* dislocation lines. The commutation rules reduce then to [18]

$$(3.25) \qquad [\mathbf{E}_3, \mathbf{E}_2] = \varepsilon\gamma\mathbf{E}_1, \qquad [\mathbf{E}_3, \mathbf{E}_1] = -\varepsilon\gamma\mathbf{E}_2, \qquad [\mathbf{E}_2, \mathbf{E}_1] = \mathbf{0},$$

and we have

$$(3.26) \qquad \rho b_g = |\gamma|.$$

If additionally

$$(3.27) \qquad \varepsilon\gamma = \text{const.} > 0,$$

then it is the case of a *flat material space* $B_g = (B, \mathbf{g})$ considered in [5].

If

$$(3.28) \qquad \gamma = 0, \qquad \mathbf{t} = t_g \mathbf{E}_3, \qquad t_g > 0, \qquad \varphi = \pi/2,$$

then it follows from Eqs. (3.16), (3.18), and (3.19) that

$$(3.29) \qquad \mathbf{l}_1 = \mathbf{E}_2, \qquad \mathbf{l}_2 = -\mathbf{E}_1, \qquad \mathbf{l}_3 = \mathbf{E}_3,$$

and (cf. Eq. (3.22))

$$(3.30) \qquad \rho \mathbf{b}_\alpha = -\mu \mathbf{E}_\alpha, \qquad \mu = t_g/2, \qquad \alpha = 1, 2.$$

In this case Eq. (3.23) with

$$(3.31) \qquad \rho b_g = \mu,$$

holds. It follows from Eqs. (3.29) and (3.30) that it is the case of principal effective *edge* dislocation lines. The commutation rules are given then by [18]:

(3.32) $\qquad [\mathbf{E}_1, \mathbf{E}_2] = \mathbf{0}, \qquad [\mathbf{E}_3, \mathbf{E}_\alpha] = \mu_\varepsilon \mathbf{E}_\alpha, \qquad \mu_\varepsilon = \varepsilon\mu, \qquad \alpha = 1, 2.$

If additionally

(3.33) $\qquad \rho = \text{const.}, \qquad \mu = \text{const.} > 0, \qquad \varepsilon = 1,$

then we obtain the case of *uniformly dense distribution* [5] of principal effective *edge* dislocations of Bianchi-type V considered in [18].

Note that if

(3.34) $\qquad \boldsymbol{\gamma} = \mathbf{0}, \qquad t_g = \|\mathbf{t}\|_g > 0,$

then it follows from Eqs. (2.22)-(2.24) that all smooth curves, except those tangent to the direction of $\mathbf{t}$ for which $b_g = 0$, can be interpreted as effective edge dislocation lines for which

(3.35) $\qquad \begin{aligned} \rho\mathbf{b} &= \mu\mathbf{m}, \qquad \mathbf{l}\cdot\mathbf{m} = 0, \\ \mu &= (t_g/2)\sin\varphi_{\mathbf{l},\mathbf{t}} > 0. \end{aligned}$

Therefore, it is a continuous distribution of edge dislocations such that

(3.36) $\qquad \rho b_g = (t_g/2)\sin\varphi_{\mathbf{l},\mathbf{t}}, \qquad 0 < \varphi_{\mathbf{l},\mathbf{t}} < \pi.$

## 4. Volterra and Frenet moving frames

Let $C[\mathbf{l}]$ denote a congruence of *mixed* effective dislocation lines located in the material space $B_g = (B, \mathbf{g})$ associated with the Bravais moving frame $\Phi = (\mathbf{E}_a)$ (Section 2). The local Burgers vector $\mathbf{b}$ of the congruence is given by Eqs. (2.22)-(2.26) with

(4.1) $$\varphi_{\mathbf{b},\mathbf{l}} \in (0, \pi).$$

Note that for effective *screw* dislocation lines $\varphi_{\mathbf{b},\mathbf{l}} = 0$ or $\pi$ and thus, according to Eqs. (2.28), (3.1) and (3.2), a congruence of effective screw dislocations is *principal* iff $\mu = 0$. If the condition (4.1) is fulfilled, then the family $\pi(\mathbf{l}, \mathbf{m})$ of planes in $B_g$ spanned by the vector fields $\mathbf{l}$ and $\mathbf{m}$ constitute an uniquely defined *two-dimensional distribution* (see [5], Appendix) and the unit vector field $\mathbf{n}$ normal to these planes is uniquely defined up to its orientation. It follows that the planes of this distribution are *local slip planes* for the congruence $C[\mathbf{l}]$ iff

(4.2) $$\mathbf{b} \cdot \mathbf{n} = 0,$$

or, equivalently:

(4.3) $$\mathbf{n}\gamma\mathbf{l} = 0.$$

The condition (4.2) means that the congruence consists of *locally Volterra-type* effective dislocation lines (cf. Section 1). For example, it is the case of a *principal congruence* of effective dislocation lines defined by the conditions (3.1) and (3.16).

The ordered triple $\Upsilon = (\mathbf{l}, \mathbf{m}, \mathbf{n})$, defined by Eqs. (4.1)-(4.3) and called further on a *Volterra moving frame*, defines the two-dimensional *oriented distribution* $\pi_{\mathbf{n}}(\mathbf{l}, \mathbf{m})$ of local slip planes associated with the considered congruence of mixed effective dislocation lines. Note that, according to Eq. (2.15), the dislocation lines of opposite orientation have local slips in opposite directions and thus these are physical opposites (see e.g [2]). However, the transformation $(\mathbf{l}, \mathbf{m}, \mathbf{n}) \to (-\mathbf{l}, -\mathbf{m}, \mathbf{n})$ of the ordered triple preserves its orientation and thus the orientation of the considered distribution is preserved. If this oriented two-dimensional distribution is *integrable* ([14], [19]; see also [5]), then through each point of $B_g$ passes an unique oriented maximal integral manifold of the distribution. These integral manifolds are virtually *slip surfaces* (Section 1) for (effective) mixed dislocations of the considered congruence $C[\mathbf{l}]$ and the unit vector field $\mathbf{n}$ defines the congruence $C[\mathbf{n}]$ of curves normal to this family of (virtual) slip surfaces. Note that for each $p \in B$ the triple $\Upsilon_p = (\mathbf{l}_p, \mathbf{m}_p, \mathbf{n}_p)$ spans the vector space $T_p(B)$ tangent to $B$ at $p$. Thus, the Bravais moving frame $\Phi$ as well as the Volterra moving frame $\Upsilon$ span the linear module $W(B)$ of all smooth vector fields on $B$ tangent to $B$ (see [5], Appendix).

Let us consider an **g**-orthonormal basis $\Im = (\mathbf{e}_a; a=1,2,3)$ of $W(B)$ being a *Frenet moving frame* (Appendix). For example, in the case of the congruence $C[\mathbf{l}]$ of mixed effective dislocation lines endowed with the local Burgers vector defined by Eqs. (4.1), (4.2) and by the condition

(4.4) $$\mathbf{e}_1 = \mathbf{l},$$

we have that $\mathbf{e}_1$ is the (unit) *tangent*, $\mathbf{e}_2$ is the *principal normal*, and $\mathbf{e}_3$ is the *second normal* of this congruence. The vector

(4.5) $$\boldsymbol{\kappa} = \kappa \mathbf{e}_2,$$

is the *curvature vector* of the congruence and the scalars $\kappa$ and $\tau$ of Eq. (A.13) are the *curvature* and *torsion* of the congruence, respectively.

Let us consider, as an example, the Frenet moving frame for a congruence of helical dislocations consisting of *cylindrical helices* ([17] and Section 1) defined by the condition that

(4.6) $$\tau = c\kappa, \quad c = \text{const.} \geq 0.$$

It follows from Eqs. (A.13), (4.4) and (4.6) that the following conditions would be fulfilled:

$$\mathbf{N} = c\mathbf{l} + \mathbf{e}_3, \qquad \nabla_{\mathbf{l}}^g \mathbf{N} = 0,$$
(4.7)
$$a^2 = \|\mathbf{N}\|_g^2 = 1 + c^2 = \text{const.}$$

So, the unit tangent $\mathbf{l}$ is inclined at the constant angle $\varphi_0$ to the vector field $\mathbf{N}$:

(4.8)
$$\cos\varphi_0 = \frac{\mathbf{N} \cdot \mathbf{l}}{a} = \frac{c}{a}, \qquad 0 \leq \varphi_0 \leq \pi/2.$$

It can be shown [17] that if

(4.9)
$$\mathbf{e}_1 = \mathbf{l} = \frac{1}{a}(\sin a\theta \mathbf{E}_1 - \cos a\theta \mathbf{E}_2 + c\mathbf{E}_3),$$
$$\mathbf{e}_2 = \cos a\theta \mathbf{E}_1 + \sin a\theta \mathbf{E}_2, \qquad \mathbf{N} = a\mathbf{E}_3,$$

the considered Bravais moving frame is parallel along the curves of the congruence $C[\mathbf{l}]$:

(4.10)
$$\nabla_{\mathbf{l}}^g \mathbf{E}_3 = 0, \qquad a = 1, 2, 3,$$

and the curvature $\kappa$ of the congruence has the form:

(4.11)
$$\kappa = \partial_1 \theta > 0,$$

then Eqs. (A.13), (4.6) and (4.9) are satisfied with

(4.12)
$$\mathbf{e}_3 = -\frac{c}{a}(\sin a\theta \mathbf{E}_1 - \cos a\theta \mathbf{E}_2) + \frac{1}{a}\mathbf{E}_3.$$

## 5. Foliation

Let $\Phi = (\mathbf{E}_a; a=1,2,3)$ be a Bravais moving frame and let $\pi = \pi(\mathbf{E}_1, \mathbf{E}_2)$ denote a family of local crystal planes spanned by the base vector fields $\mathbf{E}_1$ and $\mathbf{E}_2$. We will assume that the Riemannian material space $B_g = (B, \mathbf{g})$, where $\mathbf{g} = \mathbf{g}[\Phi]$ is given by Eqs. (2.2)-(2.4), is *foliated* by such two-dimensional distribution of local planes [5].

It means that through each point $p \in B_g$ there passes an unique maximal integral manifold of $\pi$ of the form

(5.1) $$\Sigma_c = \varphi^{-1}(c), \qquad d\varphi \neq 0,$$

where $c \in \mathbb{R}$ is a constant and $\varphi \in C^\infty(B)$. It can be shown that for each $p \in B$ there exists then a coordinate system $X = (X^A): U \to \mathbb{R}^3$, such that $p \in U$ and $X^3 = \varphi$. For any such coordinates $\partial_\alpha = \partial/\partial X^\alpha$, $\alpha = 1, 2$, is a local basis for the family $\Sigma = \{\Sigma_c, c \in \mathbb{R}\}$ of these integral manifolds and the slices

(5.2) $$\Sigma_c = \{q \in U: X^3(q) = c\}$$

belong to $\Sigma$ [5]. Thus, we can consider, at least locally, a distribution of local crystal planes with integral manifolds defined by a *distinguished coordinate system* on $B$.

The foliation of $B_g$ by $\pi = \pi(\mathbf{E}_1, \mathbf{E}_2)$ is equivalent to the condition that there are $C^\infty$-functions $C^\kappa_{\alpha\beta}$, $\alpha, \beta, \kappa = 1, 2$, on $B_g$ such that [5]:

(5.3) $$[\mathbf{E}_\alpha, \mathbf{E}_\beta] = C^\kappa_{\alpha\beta} \mathbf{E}_\kappa.$$

We will assume that additionally, the following condition is fulfilled:

(5.4) $$[\mathbf{E}_3, \mathbf{E}_\alpha] = H \mathbf{E}_\alpha, \qquad \alpha = 1, 2,$$

where $H$ is a $C^\infty$- function on $B_g$. For example, if

(5.5) $$\mathbf{E}_\alpha(p(X)) \equiv \mathbf{E}_\alpha(X^C(p)) \doteq E(X^C(p)) \mathbf{a}_\alpha(X^\kappa(p)), \qquad \mathbf{E}_3 \doteq \partial_3,$$

where, for the simplicity of notations, the vector fields $\mathbf{E}_a | U$ and $\mathbf{E}_a \circ X^{-1}$ are identified, then the condition (5.4) with

(5.6) $$H = \partial_3 \ln E$$

is fulfilled. The material metric tensor $\mathbf{g}$ takes then the so-called *geodesic form* [10] discussed in [5]:

(5.7) $$\mathbf{g}(X) \doteq g_{\alpha\beta}(X^C) dX^\alpha \otimes dX^\beta + dX^3 \otimes dX^3,$$

where the coefficients $g_{\alpha\beta}$ have the following particular form:

(5.8) $$g_{\alpha\beta}(X^A) \doteq \Psi(X^A) a_{\alpha\beta}(X^\kappa), \qquad \Psi = E^{-2}.$$

The hypersurfaces $\Sigma_c$, $c \in \mathbb{R}$, are said then to be *geodesically parallel* to the hypersurface $\Sigma_0$ and their metric tensors $\mathbf{a}_c$ are given respectively by:

(5.9) $$\begin{aligned} \mathbf{a}_c(X^\kappa) &= a_{c,\alpha\beta}(X^\kappa) dX^\alpha \otimes dX^\beta, \\ a_{c,\alpha\beta}(X^\kappa) &\doteq \Psi(X^\kappa, c) a_{\alpha\beta}(X^\kappa). \end{aligned}$$

In this case, the crystal surfaces $\Sigma_c$, $c \in \mathbb{R}$, are *umbilical* and their *mean curvatures* $H_c$ has, according to Eqs. (5.6) and (5.8), the form [10]

(5.10) $$H_c(X^\kappa) = H(X^\kappa, c),$$

where definition of the mean curvature according to SCHOUTEN [20] in place of the definition of EISENHART [10] was taken into account. The umbilical crystal surfaces have been discussed in [5].

Since, according to Eqs. (5.3) and (5.4), we have:

(5.11) $$C^3_{\alpha\beta} = 0, \qquad \alpha, \beta = 1, 2,$$

and

(5.12) $$C^1_{3\alpha} = -C^1_{\alpha 3} = H, \qquad C^2_{3\alpha} = -C^2_{\alpha 3} = H, \qquad C^3_{3\alpha} = 0, \qquad \alpha = 1, 2,$$

the covariant components $t_a$ of the vector field $\mathbf{t}$ (see Eq. (2.21)) defined by Eq. (2.19), takes the form

(5.13) $$t_1 = \varepsilon C^2_{12}, \qquad t_2 = \varepsilon C^1_{21}, \qquad t_3 = 2\varepsilon H.$$

Thus, according to Eqs. (2.5), (2.10), and (2.18), the components $\alpha^{ab}$ and $\gamma^{ab}$ of the dislocation density tensor and its symmetric part constitute the following matrices:

(5.14)
$$\left(\alpha^{ab};\ ^{a\downarrow 1,2,3}_{b\rightarrow 1,2,3}\right)=\begin{pmatrix} 0 & t_3/2 & 0 \\ -t_3/2 & 0 & 0 \\ t_2 & -t_1 & 0 \end{pmatrix},$$

and

(5.15)
$$\left(\gamma^{ab};\ ^{a\downarrow 1,2,3}_{b\rightarrow 1,2,3}\right)=\begin{pmatrix} 0 & 0 & \alpha \\ 0 & 0 & \beta \\ \alpha & \beta & 0 \end{pmatrix},$$

where

(5.16)
$$\alpha = t_2/2, \qquad \beta = -t_1/2.$$

The eigenvectors $\mathbf{\gamma}_a$ of the symmetric tensor field $\mathbf{\gamma}$ of Eqs. (2.21) and (3.3) computed with respect to the intrinsic metric tensor $\mathbf{g}$ of Eq. (2.4) are defined by

(5.17)
$$\begin{aligned}\mathbf{\gamma}\mathbf{\gamma}_a &= \gamma_a \mathbf{\gamma}_a, \\ \mathbf{\gamma}_a \cdot \mathbf{\gamma}_b &= \delta_{ab}, \qquad a,b=1,2,3,\end{aligned}$$

where the eigenvalues $\gamma_a$ of $\mathbf{\gamma}$ are roots of the determinant equation

(5.18)
$$\det\left(\gamma^{ab}-\lambda\delta^{ab}\right)=\lambda\left(\lambda^2-\gamma^2\right)=0, \qquad |\gamma|=\sqrt{\alpha^2+\beta^2}\geq 0.$$

Introducing the angle $\varphi$ by

(5.19)
$$\varphi = \arctg\left(-\frac{\alpha}{\beta}\right), \qquad \frac{\alpha}{\beta}=-\frac{t_2}{t_1},$$

we can rewrite Eq. (5.16) in the form

(5.20)
$$t_1 = -2\gamma\cos\varphi, \qquad t_2 = -2\gamma\sin\varphi.$$

A straightforward computation shows that the eigenvectors have the form:

(5.21)
$$\begin{aligned}\mathbf{\gamma}_1 &= \frac{1}{\sqrt{2}}(\mathbf{k}+\mathbf{E}_3), & \mathbf{\gamma}_2 &= \frac{1}{\sqrt{2}}(\mathbf{k}-\mathbf{E}_3), \\ \mathbf{\gamma}_3 &= \cos\varphi\,\mathbf{E}_1+\sin\varphi\,\mathbf{E}_2, & \mathbf{k} &= \sin\varphi\,\mathbf{E}_1-\cos\varphi\,\mathbf{E}_2,\end{aligned}$$

and the corresponding eigenvalues are given by:

(5.22)
$$-\gamma_1 = \gamma_2 = \gamma, \qquad \gamma_3 = 0.$$

Thus, we obtain

(5.23) $$\boldsymbol{\gamma} = \gamma\left(-\boldsymbol{\gamma}_1 \otimes \boldsymbol{\gamma}_1 + \boldsymbol{\gamma}_2 \otimes \boldsymbol{\gamma}_2\right), \qquad \gamma \geq 0,$$

and, according to Eqs. (5.13), (5.20) and (5.21), the vector field **t** takes the form

(5.24) $$\begin{aligned}\mathbf{t} &= t^a \mathbf{E}_a = 2\left(-\gamma \boldsymbol{\gamma}_3 + H \mathbf{E}_3\right), \\ t_g &= \|\mathbf{t}\|_g = 2\sqrt{\gamma^2 + H^2} > 0.\end{aligned}$$

It follows from Eqs. (5.21)-(5.24) that

(5.25) $$\boldsymbol{\gamma}\mathbf{t} = -2\gamma H \mathbf{k}, \qquad \mathbf{k} \cdot \mathbf{t} = 0,$$

what means that

(5.26) $$\boldsymbol{\gamma}\mathbf{t} = \mathbf{0} \quad \text{iff} \quad \gamma H = 0.$$

The local Burgers vector **b** of the congruence $C[\mathbf{l}]$ defined by Eqs. (2.21)-(2.26) and (5.21)-(5.24) is given by

(5.27) $$\begin{aligned}\rho \mathbf{b} &= -\gamma\left(\cos\varphi_{\mathbf{l},\mathbf{E}_3}\mathbf{k} + \cos\varphi_{\mathbf{k},\mathbf{l}}\mathbf{E}_3\right) + \mu\mathbf{m}, \\ \mu &= \frac{1}{2} t_g \sin\varphi_{\mathbf{l},\mathbf{t}}, \quad \mathbf{l}\cdot\mathbf{m} = \mathbf{t}\cdot\mathbf{m} = 0, \quad \|\mathbf{l}\|_g = \|\mathbf{m}\|_g = 1.\end{aligned}$$

It follows from Eq. (5.27), that

(5.28) $$\cos\varphi_{\mathbf{b},\mathbf{l}} = -\frac{2\gamma}{\rho b_g}\cos\varphi_{\mathbf{l},\mathbf{k}}\cos\varphi_{\mathbf{l},\mathbf{E}_3},$$

where

(5.29) $$\cos\varphi_{\mathbf{b},\mathbf{l}} = \begin{cases} 0 & \text{for edge congruences} \\ \pm 1 & \text{for screw congruences} \end{cases}.$$

For example, if

(5.30) $$\mathbf{l}\cdot\mathbf{E}_3 = 0,$$

then the congruence $C[\mathbf{l}]$ consists of effective edge dislocations with the local Burgers vector of the form:

(5.31) $$\begin{aligned}\rho \mathbf{b} &= -\gamma \cos\varphi_{\mathbf{l},\mathbf{k}}\mathbf{E}_3 + \boldsymbol{\mu}, \\ \boldsymbol{\mu} &= \mu\mathbf{m}, \qquad \mu = \sin\varphi_{\mathbf{l},\mathbf{t}}\sqrt{H^2 + \gamma^2} \geq 0.\end{aligned}$$

Note that if

(5.32) $$\gamma = 0,$$

then Eqs. (3.2), (3.35) and (3.36) with $\mu = H > 0$ are valid for any congruence $C[\mathbf{l}]$ of effective dislocation lines such that $\mathbf{l} \neq \pm \mathbf{E}_3$. Particularly, if $\varphi_{\mathbf{l},\mathbf{t}} = \varphi_{\mathbf{l},\mathbf{E}_3} = \pi/2$ (that is Eq. (5.30) holds), then

(5.33) $$\rho b_g = H,$$

and this case can be interpreted as the one defining a congruence of very small prismatic edge dislocation loops (Section 1 and e.g. [2]) normal to the direction $\mathbf{m}$ and replaced, in the continuous limit, with the infinitesimal ones [7].

Let $\Upsilon = (\mathbf{l}, \mathbf{m}, \mathbf{n})$ be a Volterra moving frame associated with $\Phi$ (Section 4). It means that the congruence $C[\mathbf{l}]$ consists of Volterra-type effective dislocation lines defined by Eqs. (2.5), (2.10), (2.18)-(2.26), and (4.1)-(4.4). The condition (5.30) means then that, up to the orientation of the vector field $\mathbf{n}$, should be

(5.34) $$\mathbf{n} = \mathbf{E}_3.$$

It is equivalent to the condition that the family $\pi_\mathbf{n}(\mathbf{l}, \mathbf{m})$ of local oriented slip planes defined by $\Upsilon$ (Section 4) is in coincidence with the distribution $\pi = \pi(\mathbf{E}_1, \mathbf{E}_2)$ of local crystal planes. In this case the considered crystal surfaces (being integral manifods of the distribution $\pi$) are virtual *glide surfaces* (see Section 1) for the considered congruence $C[\mathbf{l}]$ of Volterra-type effective dislocation lines specified by the additional conditions (5.21)-(5.24), (5.30) and (5.34). The local Burgers vector $\mathbf{b}$ of this congruence of Volterra-type effective dislocation lines is given by Eqs.(5.31) and (5.34). In this case

(5.35) $$\rho b_g = \sqrt{\gamma^2 \cos^2 \varphi_{\mathbf{k},\mathbf{l}} + (H^2 + \gamma^2) \sin^2 \varphi_{\mathbf{l},\mathbf{t}}},$$

and the congruence consists of such *edge* Volterra-type effective dislocation lines for which the *glide* (parallel to **m**-direction) as well as the *climb* (parallel to the **n**-direction) components of **b** are admitted (see Section 1). For example, if

(5.36)
$$\mathbf{l} = \mathbf{\gamma}_3,$$

then the condition (5.30) is fulfilled and we are dealing with the glide phenomenon only defined, up to the local Burgers vector orientation, by:

(5.37)
$$\rho \mathbf{b} = \mu \mathbf{m}, \quad \mathbf{m} = \mathbf{k}, \quad \mu = \sqrt{H^2 + \gamma^2}.$$

Thus, the Volterra moving frame $\Upsilon = (\mathbf{l}, \mathbf{m}, \mathbf{n})$ defined by Eqs. (5.34), (5.36) and (5.37) can be identified as the one that describes a *principal congruence* of effective edge dislocation lines (Section 3) located on glide surfaces and accompanied with a distribution of secondary point defects.

### 6. Final remarks

Let us consider the congruence $C[\mathbf{l}]$ of cylindrical helices (see Section 1 and the example discussed in Section 4) defined by the condition that the angles $\theta$ and $\varphi$ of Eqs. (4.11) and (5.19), respectively, are related by

(6.1)
$$\varphi = a\theta, \quad a > 0.$$

It follows from Eqs. (4.11), (5.21), (5.24), and (5.27) that

(6.2)
$$\mathbf{l} = \frac{1}{a}(\mathbf{k} + c\mathbf{E}_3),$$

and the local Burgers vector of the congruence is given by

(6.3)
$$\rho \mathbf{b} = -\frac{\gamma}{a}(c\mathbf{k} + \mathbf{E}_3) + \sqrt{H^2 + \gamma^2}\,\mathbf{m}.$$

Note that if the torsion of the congruence $C[\mathbf{l}]$ equals zero, that is

(6.4) $$c = 0, \quad a = 1,$$

then, according to Eqs. (4.7), (4.9), (4.12) and (5.21), we have

(6.5) $$\mathbf{e}_1 = \mathbf{l} = \mathbf{k}, \quad \mathbf{e}_2 = \mathbf{\gamma}_3, \quad \mathbf{e}_3 = \mathbf{E}_3.$$

For example, if

(6.6) $$\Upsilon = (\mathbf{l}, \mathbf{m}, \mathbf{n}) = (\mathbf{e}_1, -\mathbf{e}_3, \mathbf{e}_2),$$

then

(6.7) $$\rho \mathbf{b} = -\left(\gamma + \sqrt{H^2 + \gamma^2}\right)\mathbf{E}_3.$$

So, we have defined a congruence of effective Volterra-type edge dislocations of torsion zero, located on crystal surfaces with tangent planes being osculating planes (that is $\pi(\mathbf{E}_1, \mathbf{E}_2) = \pi(\mathbf{e}_1, \mathbf{e}_2)$) normal to the local Burgers vector direction. The local slip planes are identical with the rectifying planes $\pi(\mathbf{e}_1, \mathbf{e}_3)$. Thus, it is a congruence of effective dislocation lines being intersections of two orthogonal families of surfaces embedded in $B_g$: umbilical crystal surfaces (on which the dislocations are located; see [5] and Section 5) and slip surfaces (in which the dislocations can move). We will assume additionally that the crystal surfaces are *minimal varieties* in $B_g$:

(6.8) $$H = 0.$$

In this case [17]

(6.9) $$\rho \mathbf{b} = -2\gamma \mathbf{E}_3,$$

and

(6.10) $$\rho b_g = 2|\gamma|.$$

The crystal surfaces defined in this manner are *totally geodesic* surfaces and thus the vectors normal to these surfaces are parallel in the enveloping material space [10], that is

(6.11) $$\nabla^g \mathbf{m} = 0, \quad \mathbf{m} = -\mathbf{e}_3,$$

where Eq. (6.6) was taken into account. The totally geodesic (crystal) surfaces are an evident generalization of (crystal) planes of Euclidean 3-space [10].

It is known that a dislocation of torsion zero lying in a Euclidean plane experiences a *static straightening force* per unit length of dislocation, acting against the direction of the curvature vector $\kappa$ of Eq.(4.7) and tending to straighten the line [2], that is the force $\mathbf{F}$ such that [17]

(6.12) $$\mathbf{F} = -S\mathbf{\kappa} = -S\kappa \mathbf{e}_2, \quad \mathbf{F} \cdot \mathbf{l} = \mathbf{F} \cdot \mathbf{m} = 0,$$
$$S > 0, \quad \kappa > 0; \quad [\mathbf{e}_2] = [\kappa] = \text{cm}^{-1},$$

where $\kappa$ is the curvature given by Eq. (4.11) and $\mathbf{e}_2$ is the principal normal of the considered congruence $C[\mathbf{l}]$. If Eq. (6.12) is assumed to be valid for the considered congruence of effective prismatic edge dislocation lines of zero torsion in the material space $B_g$, then

(6.13) $$F_g = \|\mathbf{F}\|_g = S\kappa, \quad [F_g] = \text{kgcm}^{-1}, \quad [S] = \text{kg}.$$

A dislocation line in a crystal will remain curved only if there is a shear stress that produces a force on the dislocation needed to maintain its curvature $\kappa$ [2]. So, let $\mathbf{T}$ be a symmetric tensor defined in the material space $B_g$ and identified with an *internal stress tensor* dependent on the distribution of dislocations and secondary point defects, and let $T$ denote the field of *resolved shear stresses* defined by

(6.14) $$T = \mathbf{mTn},$$

and acting in the oriented local slip planes $\pi_n(\mathbf{l}, \mathbf{m})$ in the direction $\mathbf{m}$ of the local Burgers vector $\mathbf{b}$. The static straightening force $\mathbf{F}$ is normal to these slip planes. We will assume, generalizing the elastic model of forces acting on single dislocation lines (see e.g. [2]), that an effective dislocation line of strength $b_g$ will be in *local equilibrium* in its curved position when

(6.15) $$T = \frac{F_g}{b_g}.$$

Substituting $F_g$ from Eq. (6.13), we obtain

(6.16) $$S = \frac{b_g}{\kappa} T.$$

The quantity $S$ of Eq. (6.16) has units of energy per unit length and thus the effective dislocation line has a *line tension* that is analogous to the surface tension of a soap bubble or liquid [2]. Note that Eq. (6.16) generalizes the expression of line tension of a curved dislocation lying in a Euclidean plane (cf. [3]).

The strength $b_g$ of the considered effective dislocation lines can be written, according to Eq. (6.10), in terms of scalar characteristics $\gamma$ and $\rho$ of the continuous distribution of dislocations:

(6.17) $$b_g = \frac{2|\gamma|}{\rho}.$$

Eqs. (6.16) and (6.17) lead to the following expression of shear stresses required to bend effective dislocation lines of the congruence and of the line tension of these dislocations:

(6.18) $$T = E_d \rho \kappa, \qquad S = E_d \rho b_g,$$

where the scalar $E_d > 0$ is defined as

(6.19) $$E_d = \frac{S}{2|\gamma|},$$

and it has units of energy, that is $[E_d] = \text{kgcm}$. We see that the considered congruence of effective prismatic edge dislocation lines can be endowed with a finite *self-energy* $E_d$ [17].

**Appendix**

Let $\gamma: I \to M$, where $I \subset \mathbb{R}$ is an interval, be a smooth *parametrized curve* in a 3-manifold $M$. The curve $\gamma$ is called *regular* if [19]

(A.1) $$\forall t \in I, \quad \dot{\gamma}(t) = \gamma_* \left(\frac{d}{dt}\right)_t \neq 0,$$

where $\dot{\gamma}(t) \in T_{\gamma(t)}(M)$ denotes the vector tangent to $\gamma(I) \subset M$ at the point $\gamma(t)$.

*Remark* [21]

Let $\chi: M \to N$ be a smooth mapping from the manifold $M$ into a manifold N and let us denote by W(M) and W(N) the linear moduli of smooth vector fields on $M$ and $N$, respectively (see [5], Appendix). The *tangent mapping* $\chi_*: W(M) \to W(N)$ is defined by

(A.2) $$\forall p \in M, \quad (\chi_* \mathbf{v})_{\chi(p)} = d_p \chi(\mathbf{v}_p),$$

where $d_p \chi: T_p M \to T_{\chi(p)} N$ is a mapping between tangent spaces acting, for any $\mathbf{v}_p \in T_p(M)$ interpreted as a linear first-order differential operator with constant coefficients (see [5], Appendix), according to the rule:

(A.3) $$\forall f \in C^\infty(N), \ d_p\chi(\mathbf{v}_p)(f) = \mathbf{v}_p(f \circ \chi).$$

If $\mathbf{v} \in W(M)$, then we define the vector field $\chi_*\mathbf{v} \in W(N)$ by

(A.4) $$\chi_*\mathbf{v} = d\chi \circ \mathbf{v} \circ \chi^{-1}.$$

If $\mathbf{v} \in W(M)$, then a smooth parametrized curve $\gamma : I \to M$ such that

(A.5) $$\forall t \in I, \ \mathbf{v}_{\gamma(t)} = \gamma_*\left(\frac{d}{dt}\right)_t,$$

is called an *integral curve* of the vector field $\mathbf{v}$ [1]. The vector field $\mathbf{v}_\gamma : I \to W(M)$ defined as $\mathbf{v}_\gamma(t) = \mathbf{v}_{\gamma(t)}$ is then a field of vectors tangent to the integral curve $\gamma$.

*Theorem* 1 [22]

If $\mathbf{v} \in W(M)$, then for every $p \in M$, there exists an integral curve of $\gamma = \gamma_p^\mathbf{v}$ of $\mathbf{v}$ such that

(1) $\gamma_p^\mathbf{v} : I_p = (a_p, b_p) \to M, \ 0 \in I_p; \ \gamma_p^\mathbf{v}(0) = p.$

(2) Every integral curve $\sigma$ of $\mathbf{v}$ such that $\sigma(0) = p$ is a restriction of the curve $\gamma_p^\mathbf{v}$ to an open interval $I \subset I_p, \ 0 \in I$.

The curve $\gamma = \gamma_p^\mathbf{v}$ of *Theorem* 1 is called a *maximal integral curve* of $\mathbf{v}$ passing through the point $p$. If every maximal integral curve of a vector field $\mathbf{v} \in W(M)$ is defined on the interval $I = \mathbb{R}$, then $\mathbf{v}$ is called a *complete* vector field on $M$.

The parametrized curves $\gamma : I \to M$ and $\sigma : I' \to M$ are called *equivalent* if there exists a diffeomorphism $\varphi$ of the open interval $I$ onto the open interval $I'$ such that $\sigma = \gamma \circ \varphi$ and $d\varphi \neq 0$. Then

(A.6) $$\sigma_*\left(\frac{d}{dt'}\right)_{t'=\varphi(t)} = \dot\varphi(t)\gamma_*\left(\frac{d}{dt}\right)_t, \quad \dot\varphi = \frac{d\varphi}{dt}.$$

*Theorem* 2 [19], [23]

If the parametrized curves $\gamma$, $\sigma$ are eqivalent and one of them is regular, then the second curve is also regular and

(A.7) $$C = \gamma(I) = \sigma(I') \subset M.$$

The class $[\gamma]$ of all regular parametrized curves equivalent to $\gamma$ is called a *geometric curve* and can be identified with the set $C$ of the manifold $M$ [23]. If additionally for each curve $\sigma \in [\gamma]$ we have $\sigma = \gamma \circ \varphi$ and $d\varphi > 0$, then the curve is called *oriented* (parametrized curves representing an oriented geometric curve define the same order of points of the set $C$ of Eq. (A.7)) A set of all geometric curves defined by a nonvanishing everywhere vector field $\mathbf{v} \in W(M)$ is called a *congruence*.

Let $\gamma : [a, b] \to M$ be a regular parametrized curve in a Riemannian manifold $M_g = (M, \mathbf{g})$. The length $l(\gamma)$ of $\gamma$ is defined as

(A.8) $$l(\gamma) = \int_a^b \|\dot{\boldsymbol{\gamma}}(t)\|_g \, dt,$$

where $\dot{\boldsymbol{\gamma}}(t) \in T_{\gamma(t)}(M)$ is defined by Eq. (A.1) and $\|\mathbf{v}_p\|_g$ denotes the norm of a vector $\mathbf{v}_p \in T_p M$ tangent to the manifold $M_g$ at the point $p \in M$. Since $l(\gamma) = l(\sigma)$ for all equivalent parametrized curves, so the number $l(\gamma)$ is well-defined *length of the geometric curve* $C_\gamma = \gamma(I) \subset M_g$ identified with the class $[\gamma]$ of equivalent curves. If $\gamma$ is a regular integral curve of the vector field $\mathbf{v}$ on $M_g$, then

(A.9) $$l(\gamma) = \int_a^b \|\mathbf{v}_{\gamma(t)}\|_g \, dt.$$

Particularly, if $\gamma = \gamma_p^{\mathbf{v}}: I_p \to M$, $I_p \subset \mathbb{R}$, is the maximal integral curve of **v** passing through $p \in M$, then the length of the geometric curve $C_p^m(\mathbf{v}) = \left[\gamma_p^{\mathbf{v}}\right]$ is defined as

(A.10) $$l_p^m(\mathbf{v}) = l(\gamma_p^{\mathbf{v}}),$$

It follows that for each $p \in M$ there exists one and only one *maximal geometric curve* $C_p^m(\mathbf{v})$ passing through this point.

Let $\gamma: I \to M_g$ be a regular parametrized curve of the length $l(\gamma)$. The curve $\gamma$ is called *normalized* [19] or *natural* [21] if

(A.11) $$\forall t \in I \quad \|\dot{\boldsymbol{\gamma}}(t)\|_g = 1.$$

Every regular curve is equivalent to a normalized curve [22]. This equivalence defines a diffeomorphism $\varphi: I \to (0, l(\gamma))$ defined by

(A.12) $$\varphi(t) = \int_a^t \|\dot{\boldsymbol{\gamma}}(r)\|_g \, dr, \quad t \in I = (a, b),$$

and called the change of a parameter $t \in I$ on the *natural parameter* $s = \varphi(t)$, $0 \leq s \leq l(\gamma)$. The curve $\sigma = \gamma \circ \varphi^{-1}: (0, l(\gamma)) \to M_g$ is a normalized curve equivalent to the curve $\gamma$ and called its *natural parametrization*.

Let **v** be a nonvanishing vector field tangent to the three-dimensional Riemannian manifold $M_g = (M, \mathbf{g})$ and let us denote by $C[\mathbf{v}]$ the *congrueunce* defined by **v**, i.e. the set of all *geometric curves* defined by **v**. The vector field $\mathbf{l} = \mathbf{v}/\|\mathbf{v}\|_g$ defines unit vectors tangent to geometric curves of the congruence. We will generalize the the Frenet formulae for integral curves of **v** (see. e.g. [13], [20], [24] and [25]) assuming the existence (at least locally) of an orthonormal base $\{\mathbf{e}_a; \mathrm{a} = 1, 2, 3\}$ of vector fields on $M_g$ such that $\mathbf{e}_1 = \mathbf{l}$ and the following *generalized Frenet formulae* are valid [17]:

(A.13)
$$\begin{aligned}\boldsymbol{\kappa} &= \nabla^g_{\mathbf{e}_1}\mathbf{e}_1 = \kappa\mathbf{e}_2, & \kappa \rangle 0, \\ \nabla^g_{\mathbf{e}_1}\mathbf{e}_2 &= -\kappa\mathbf{e}_1 + \tau\mathbf{e}_3, & \tau \geq 0, \\ \nabla^g_{\mathbf{e}_1}\mathbf{e}_3 &= -\tau\mathbf{e}_2, & \kappa, \tau \in C^\infty(M),\end{aligned}$$

where $\nabla^g$ denotes the Levi-Civita covariant derivative based on the Riemannian metric **g** (e.g. [23]). The basis $\Upsilon = (\mathbf{e}_a, a = 1, 2, 3)$ of W(*M*) consists then of *Frenet vector fields* of the congruence: $\mathbf{e}_1$ is the *tangent*, $\mathbf{e}_2$ is the (*principal*) *normal*, and $\mathbf{e}_3$ is the *binormal* (called also the second normal) of the congruence and will be called a *Frenet moving frame*. The vector field $\boldsymbol{\kappa}$ is the curvature vector of the congruence. The scalars $\kappa$ and $\tau$ denote the curvature and torsion of the congruence, respectively. A Frenet moving frame defines (at least locally) three two-dimensional distributions of planes (see [5], Appendix): $\pi(\mathbf{e}_1, \mathbf{e}_2)$-*osculating planes*, $\pi(\mathbf{e}_2, \mathbf{e}_3)$-*normal planes* and $\pi(\mathbf{e}_3, \mathbf{e}_1)$-*rectifying planes*.